\documentclass[onecolumn,manuscript,preprint]{revtex4}
\usepackage{dcolumn}
\usepackage{amsmath}
\usepackage[dvips]{epsfig}      % to include PostScript figures
\usepackage{picture}
\makeatletter
\def\btt#1{\texttt{\@backslashchar#1}}%
\DeclareRobustCommand\bblash{\btt{\@backslashchar}}%
\makeatother

\begin{document}
\draft
\title{\bf  Organometallic Wires Constructed from Transitional Metals and Anthracene: A Theoretical Study}
\author{Xianlong Wang, Xiaohong Zheng and Zhi Zeng$\footnote{Correspondence author: zzeng@theory.issp.ac.cn}$}
\affiliation {Key Laboratory of Materials Physics, Institute of Solid State Physics, Chinese Academy
of Sciences, Hefei 230031, P.R. China }
%\affiliation{Graduate School of the Chinese Academy of Sciences}%Lines break automatically or can be forced with \\
\date{\today}% It is always \today, today, but you may specify any date with \date.

%----------------------------------------------abstract-----------------------

\begin{abstract}
The properties of organometallic
wires [TM$_2$(Ant)]$_{\infty}$ constructed with transitional
metals (TM = Sc, Ti, V, Cr, Mn and Fe) and anthracene (Ant) are
investigated by first-principles calculations. As the gap between HOMO
(Highest Occupied Molecular Orbital) and LUMO (Lowest Unoccupied Molecular
Orbital) of Ant is much smaller than that of benzene (Bz),
much larger charge transfer (CT) occurs between TMs and Ant, which results in
much more diverse magnetic properties in [TM$_2$(Ant)]$_{\infty}$ than in [TM(Bz)]$_{\infty}$.
Particularly, [V$_2$(Ant)]$_{\infty}$ and [Cr$_2$(Ant)]$_{\infty}$ are found
to be half-metallic ferromagnets. As a result of this and the better structural stability, compared with
[TM(Bz)]$_{\infty}$, [TM$_2$(Ant)]$_{\infty}$ (like [V$_2$(Ant)]$_{\infty}$ and [Cr$_2$(Ant)]$_{\infty}$)
may be better candidates of spintronic devices. Furthermore, as the HOMO-LUMO gap of small pieces
of graphene (SPG), such as pentacene and coronene, decreases with
the increase of polycyclic number, the CT effects may also fit for the TM-SPG
 sandwich polymers which can also act as good spintronic materials.

\end{abstract}

%----------------------------------------------end abstract-------------------

%\pacs{73.40.Cg}% PACS, the Physics and Astronomy
%\pacs{75.50.Pp, 71.22.+i, 71.15.Nc}
 \maketitle

%-----------------------------------------Introduction------------------------

\section{\bf Introduction}

One-dimensional organometallic sandwich polymers (OSPs) have
attracted wide interests in recent years due to their
potential applications in spintronic devices and information
storage \cite{ref0MiyajimaK,ref0LyonJT,ref0MiyajimaK1,
ref0WangJL,ref0MaslyukVV,ref0KoleiniM,ref0XiangHJ,ref0WengHM,ref0WengHM1,ref0NagaoS,ref0ZhouLP,
ref0ShenL,ref0KurikawaT,ref0KatzTJ,ref0WuXJ,ref0WangL}. One of the extensively
studied systems is TM$_n$Bz$_{n+1}$ which demonstrates versatile properties.
The property diversity arises from two facts.
On one hand, the properties of [TM(Bz)]$_{\infty}$ are highly dependent on the TM ions.
For example, [V(Bz)]$_{\infty}$ and [Mn(Bz)]$_{\infty}$
are ferromagnetic half-metals, while [Cr(Bz)]$_{\infty}$ is a
nonmagnetic insulator. On the other hand, the type of organic molecules
can also significantly influence the properties of the one-dimensional
OSPs. For example, no magnetic moment appears in
[Mn(Cp)]$_{\infty}$ (Cp = cyclopentadienyl), while
[Cr(Cp)]$_{\infty}$  has a ferromagnetic ground
state.\cite{ref0NagaoS,ref0ZhouLP,ref0ShenL}
It is believed that the ferromagnetism obsvered in such OSPs (like V$_n$Bz$_{n+1}$\cite{ref0MiyajimaK})
may qualify them as good materials for spin polarized transport.\cite{ref0WangJL,ref0MaslyukVV,ref0KoleiniM,ref0XiangHJ}

Because of the abundant magnetic properties and the possibility as candidates for spintronic  devices
and information storage, recently, the study on the one-dimensional OSPs has been extended to the
cases of TM-pentalene (Pn), TM-naphthalene (Np) and
TM-Ant.\cite{ref0KurikawaT,ref0KatzTJ,ref0WuXJ,ref0WangL} The sandwich like
clusters of TM-Np and TM-Ant were firstly synthesized in
1999.\cite{ref0KurikawaT} Further theoretical investigations showed
that [V$_2$(Ant)]$_{\infty}$ is a half-metallic ferromagnet.
Since in [TM$_2$(Ant)]$_{\infty}$, the Ant are connected by two TM ions, as
shown in Fig 1(a), compared with [TM(Bz)]$_{\infty}$ and [TM(Cp)]$_{\infty}$, it should have
better stability and conductivity and it may be more suitable for
spintronic devices and information storage. Note that, previous theoretical investigation
pointed out a possible structural transition from \emph{D$_{6h}$} to \emph{D$_2$} symmetry appearing on
V$_n$Bz$_{n+1}$.\cite{ref0WangJL} Consequently, systematic investigate on this kind of
polymers is quite interesting and necessary. Therefore, in this work, the
structural, electronic and magnetic properties of
[TM$_2$(Ant)]$_{\infty}$ (TM=Sc, Ti, V, Cr and Mn) are
investigated by first principles calculations. We find that
magnetic properties in  [TM$_2$(Ant)]$_{\infty}$ are much different
from those in TM(Bz)$_{\infty}$ and the difference arises from the different HOMO-LUMO
gap of the organic molecules and the subsquent different CT between the TMs
and the organic molecules. Especially, we find that [V$_2$(Ant)]$_{\infty}$ and
[Cr$_2$(Ant)]$_{\infty}$ are half metals with the 100\% spin polarized density of states (DOS)
around the Fermi level, which is of great importance for spin transport.

 \section{Compupational Details}
Full-electron density functional theory (DFT) implemented in the
Dmol$^3$ package \cite{ref0DellyB,ref0DellyB1} is used in our
calculations. The Perdew, Burke, and Ernzerhof (PBE)
\cite{ref0PerdewJP} functional is used for exchange and
correlation interaction in the generalized gradient
approximation(GGA). The basis set is composed of double numerical
basis and a polarized function (DNP) is chosen. The supercell is
chosen as 20$\times$20$\times$\emph{c} \AA$^3$, where $c$ is the lattaice
constant along the periodic direction and is varied for different TM
cases [see Table I]. The corresponding Brillouin zone is sampled by a 1$\times$1$\times$50
Monkhorst and Pack grid.\cite{ref0MonkhorstHJ} All structures are
fully relaxed.

In order to investigate the different magnetic configurations and
Peierls transition in [TM$_2$(Ant)]$_{\infty}$, a supercell
containing four TMs and two Ants are used. For each kind of
[TM$_2$(Ant)]$_{\infty}$, as shown in Figs. 1(b)-(f), five magnetic
configurations are considered. The combination of two characters \emph{XY} (\emph{X} and
\emph{Y} correspond to \emph{N}(non-magnetic),
\emph{F}(ferromagnetic) and \emph{A}(anti-ferromagnetic)) are adopted
for representing the related magnetic configurations, with \emph{X} and
\emph{Y} describing the intralayer and interlayer magnetic coupling of TM
atoms, respectively. For example, as shown in Fig. 1(d), \emph{FA}
means the ferromagnetic coupling between the inner layer TM atoms and
the anti-ferromagnetic coupling between two nearest neighbor layers.

\section{\bf Results and Discussions}

After full relaxation, the calculated structural, electronic and magnetic
properties of [TM$_2$(Ant)]$_{\infty}$ are summarized in Table I.
Peierls transition does not occur in [TM$_2$(Ant)]$_{\infty}$,
and no magnetic moment is observed in the cases of
[Sc$_2$(Ant)]$_{\infty}$ and [Mn$_2$(Ant)]$_{\infty}$.
In contrast, Ti and Fe are spin polarized,
but the magnetic ground states of [Ti$_2$(Ant)]$_{\infty}$
and [Fe$_2$(Ant)]$_{\infty}$ are \emph{FA} and \emph{AF},
respectively. Hence, the total magnetic moments of
[Ti$_2$(Ant)]$_{\infty}$ and [Fe$_2$(Ant)]$_{\infty}$ supercells
are zero. In addition, our calculations predict that \emph{FF} is
the ground state of both [V$_2$(Ant)]$_{\infty}$ and
[Cr$_2$(Ant)]$_{\infty}$, and that the supercells of [V$_2$(Ant)]$_{\infty}$
and [Cr$_2$(Ant)]$_{\infty}$  respectively possess a magnetic moment of
2.00$\mu_{B}$ and 0.96$\mu_{B}$ per TM ion, with a finite negative
magnetic moment formed on the Ant part. Further analysis shows that
[V$_2$(Ant)]$_{\infty}$ and [Cr$_2$(Ant)]$_{\infty}$ are
half-metallic ferromagnets which can be seen from the band structures of
[V$_2$(Ant)]$_{\infty}$ and [Cr$_2$(Ant)]$_{\infty}$ shown in
Figs. 2(a) and 2(b). The [V$_2$(Ant)]$_{\infty}$ total DOS and the partial DOS of V 3\emph{d}  are
shown in Fig. 2(c). It is found that the 100\%
spin-polarized DOS around Fermi level mainly come from the V 3\emph{d}
orbitals, with small contribution from the C 2\emph{d}.
A gap of 0.76 eV and 1.25 eV for [V$_2$(Ant)]$_{\infty}$
and [Cr$_2$(Ant)]$_{\infty}$ respectively is observed in the spin majority
channel.  Thus, [V$_2$(Ant)]$_{\infty}$ and [Cr$_2$(Ant)]$_{\infty}$ may be good
candidates for spintronic devices and information storage.
Note that our relsults on [V$_2$(Ant)]$_{\infty}$ presented here agree well with former
calculations.\cite{ref0WangL}

In both [TM$_2$(Ant)]$_{\infty}$ and [TM(Bz)]$_{\infty}$, TMs are located in a crystal field with
\emph{D$_{6h}$} symmetry. Under \emph{D$_{6h}$} symmetry crystal
field, as shown in Fig. 3(a), TMs \emph{d} orbital splits into a
a$_1$ (d$_{3z^2-r^2}$), 2-fold e$_2$ (d$_{xy}$ and d$_{x^2-y^2}$)
and e$_1$ (d$_{xz}$ and d$_{yz}$) orbitals. In the meantime, both Ant and Bz containing 14 and 6 C atoms
obey the H\"ukckel rule, which means that a stable aromatic
molecule should have 4\emph{m}+2 carbon atoms (where \emph{m} is an integer). However, the
magnetic behaviors of [TM$_2$(Ant)]$_{\infty}$  are totally different from
those of [TM(Bz)]$_{\infty}$.\cite{ref0XiangHJ} For example,
[Mn(Bz)]$_{\infty}$ is a ferromagnet with 1.0$\mu_{B}$ magnetic
  moment per Mn ions, and non-magnetic state is  the ground state
  of [Cr(Bz)]$_{\infty}$. Whereas, this situation is reversed in
   [TM$_2$(Ant)]$_{\infty}$, since [Mn$_2$(Ant)]$_{\infty}$ do not show any
   magnetic moment while  Cr ions of
 [Cr$_2$(Ant)]$_{\infty}$ are ferromagnetically spin-polarized with 1$\mu_{B}$
 magnetic moment. Both [V(Bz)]$_{\infty}$ and [V$_2$(Ant)]$_{\infty}$ are a spin-polarized sandwich polymers,
but the magnetic moments of V ions in [V(Bz)]$_{\infty}$ and
[V$_2$(Ant)]$_{\infty}$ are approximately
1.0$\mu_{B}$\cite{ref0XiangHJ} and 2.0$\mu_{B}$, respectively.
An interesting question arises from these results: Why is there so big difference
in the magnetic properties of [TM$_2$(Ant)]$_{\infty}$ and [TM(Bz)]$_{\infty}$?

Further analysis shows that, compared with [TM(Bz)]$_{\infty}$,
there is a larger CT between TMs and Ant. We find that the magnetic properties
of [TM$_2$(Ant)]$_{\infty}$ are attributed to the CT effect which was once
proved to be responsible for the magnetic behaviors of
[TM(Cp)]$_{\infty}$.\cite{ref0ShenL} For example, in [V(Bz)]$_{\infty}$, V (3d$^3$4s$^2$)
has five valence electrons occupy the \emph{d} orbitals with two and three electrons
respectively occupying the \emph{a$_1$} and doubly degenerate
\emph{e$_2$} orbitals, because of two \emph{s} electrons fill 3\emph{d} shell\cite{ref0MaslyukVV}.
Hence, one unpaired electron exists in
\emph{e$_2$} orbitals, and [V(Bz)]$_{\infty}$ shows a magnetic moment of
1.0$\mu_{B}$ on each V ion. However, in [V$_2$(Ant)]$_{\infty}$, one
electron at V \emph{d} orbital is transferred to Ant, and
\emph{e$_2$} has two unpaired electrons. Thus, V ions in
[V$_2$(Ant)]$_{\infty}$ possess a 2.0$\mu_{B}$ magnetic moment. In order to
illustrate this effect more clearly, calculations are performed, with the unit cell containing two
TMs and one Ant and the initial magnetic coupling of two
TMs arranged as ferromagnetic. For each
[TM$_2$(Ant)]$_{\infty}$, the calculated total unit-cell magnetic moments per TM
 are shown in Fig. 3(b), in which the magnetic moments and
schematic occupation of valence electrons obtained by CT model
are also presented. We can find that our calculated results agree well
with the CT model. Furthermore,
[TM$_2$(Ant)]$_{\infty}$ and [TM(Cp)]$_{\infty}$  are quite
similar in magnetic properties, since Cp containing 5 C atoms does not obey the
H\"uckel rule, Cp tends to get one additional electron from TMs.
Therefore, CT occurs in [TM(Cp)]$_{\infty}$.\cite{ref0ShenL} However, according
to H\"uckel rule, Ant is a stable aromatic molecule. We want to inspect why
there is so big CT in [TM$_2$(Ant)]$_{\infty}$. Generally,
the key factors determining the CT in this kind of polymers are: (1)
the average number of carbon atom (ANC) per TM; (2) the strength
of frontier orbital hybridization. From [TM(Bz)]$_{\infty}$ to
[TM$_2$(Ant)]$_{\infty}$, the ANC increases from six to seven. This
small change of ANC should not induce large CT difference between
[TM(Bz)]$_{\infty}$ and [TM$_2$(Ant)]$_{\infty}$.  Then
the large CT in  [TM$_2$(Ant)]$_{\infty}$ might arise from the second factor.

Our calculations show that the HOMO-LUMO gap of Ant is 2.7
eV, which is smaller than that of Bz(5.5 eV).
Comparing with Bz, the LUMO of Ant is much closer to the HOMO of
TM ions. Thus, the frontier orbital hybridization  between TMs
   and Ant is much stronger in [TM$_2$(Ant)]$_{\infty}$ than in
   [TM(Bz)]$_{\infty}$ and the CT appears on [TM$_2$(Ant)]$_{\infty}$.
This is also illustrated clearly in Figs. 3(c) and 3(d), where the
partial DOS of TM 3\emph{d} and Ant C 2\emph{p} are shown.
Taking Fermi level as the reference, the HOMO of both Bz and Ant are
localized around -5.5 eV. However, as the gap of Ant is smaller
than that of Bz, the minimum energy of hybdridization of V 3\emph{d} and C 2\emph{p}
electrons in [V$_2$(Ant)]$_{\infty}$ is -4 eV, which is
 lower than that of [V(Bz)]$_{\infty}$ (-3 eV).
Thus, as shown in Fig 3(c) and (d), compared with [V(Bz)]$_{\infty}$, the minority electrons of V in
[V$_2$(Ant)]$_{\infty}$ is transferred to Ant, which can also be
found in [V(Cp)]$_{\infty}$.\cite{ref0ShenL} Consequently,
from [V(Bz)]$_{\infty}$ to [V$_2$(Ant)]$_{\infty}$, the magnetic
moment of V ion increases from 1$\mu_{B}$ to 2$\mu_{B}$. The
schematic view of this opinion is also shown in Fig. 3, where the
frontier orbitals (HOMO and LUMO) of Bz, V and Ant are
approximately shown in the left, middle and right, respectively.
The bigger arrow means the larger CT in [TM(Bz)]$_{\infty}$ than
that in [TM(Bz)]$_{\infty}$.

Double exchange (DE) was attributed as one of the ferromagnetic origins of the
 in [TM(Bz)]$_{\infty}$ and [TM$_2$(Pn)]$_{\infty}$.\cite{ref0XiangHJ,ref0WuXJ} However,
in [V(Bz)]$_{\infty}$, the ferromagnetic state was believed to arise from the
negative
polarization of carbon atoms since magnetic
moments anti-polarized by TMs appears on benzene.\cite{ref0WengHM}
In addition, in [V(Cp)]$_{\infty}$, the ferromagnetism originates from the CT
effects since there is CT between TMs and Cp.  \cite{ref0ShenL}
In the cases of [V$_2$(Ant)]$_{\infty}$ and [Cr$_2$(Ant)]$_{\infty}$ that are
studied in this work,
strong coupling between TMs and Ant within the energy range of -4 to
0 eV can be found [see Figs. 2(c) and 3(d)]. Furthermore, the
wave functions at the $\Gamma$ point of [V$_2$(Ant)]$_{\infty}$ and
[Cr$_2$(Ant)]$_{\infty}$ near the Fermi level also confirm the
strong TMs-Ant coupling. Therefore, by comparing CT and negative
polarization effect, it can conclude that the DE may be mainly responsible for the
ferromagnetism.

Furthermore, Ant can be considered as a small piece of graphene (SPG)
with multi-Bzs. SPG with more benzene rings, such as coronene and pentacene (Pe),
can also be used for constructing this kind of sandwich like
nanostructures and nanowires. As a matter of fact, the
sandwich like TM-corroene
polymers have already been synthesized.\cite{ref0PozniakBP,ref0FosterNR} Such kind of
polymers built with SPG and TMs may also be suitable materials for
spintronic devices and information recording. Except for the
properties of better stability and conductivity mentioned at the
beginning, as shown in Figs. 4(a) and 4(b), TM-SPG polymers can be built into
warped and branched strutures. In Fig. 4(a), the relaxed
structure of Sc$_4$V$_4$(Ant)$_5$ cluster are shown. As the radius of
Sc is larger than the that of V, Sc$_4$V$_4$(Ant)$_5$
cluster shows a warp structure. By co-doping and introducing
TM defects, warped structure can be realized in this kind of sandwich
polymers. Fig. 4(b) shows the schematic branched structure
constructed with Pe, Bz and TMs. In fact, warped and branched structures already achieved in
carbon nanotubes and nanowires\cite{ref0LiJ,ref0MengGW} are central to the design of
nanodevices and nanocircuits. While the different shapes and versatile properties exhibited by
the sandwich like TM-SPG nanostructures and nanowires may qualify them as
promising building blocks for spintronics devices and
information storage. As the HOMO-LUMO gap of SPG decrease
with the size increase of SPG, the CT effects and magnetic
behaviors in the [V$_2$(Ant)]$_{\infty}$ may also
exists in TM-SPG. Thus, further experimental and theoretical
investigations are highly suggested in this interesting field.

\section{\bf Conclusion}
In conclusion, the properties of [TM$_2$(Ant)]$_{\infty}$ are
investigated by first principles calculations. As the HOMO-LUMO gap of Ant
is smaller than that of Bz,  there is a large CT in
[TM$_2$(Ant)]$_{\infty}$, which results in different magnetic behaviors in [TM$_2$(Ant)]$_{\infty}$
from TM(Bz)$_{\infty}$.
Especially, [V$_2$(Ant)]$_{\infty}$ and [Cr$_2$(Ant)]$_{\infty}$ exhibit 100\%
spin polarized DOS around Fermi level which is of great importance for spin transport.
Furthermore, we indicate that CT effect and magnetic behaviors shown in
[V$_2$(Ant)]$_{\infty}$ may also exist in TM-SPG.
Finally, due to the excellent
stability, conductivity and unique shape (like branch and warp) that can be
possibly realized, TM-SPG may be prospective materials
for spintronics and information storage.

\noindent
\section*{\bf ACKNOWLEDGEMENTS}
This work was supported by the National Science Foundation of
China under Grant No 10774148, No 10904148 and No 10847162, the special Funds for Major State
Basic Research Project of China(973), Knowledge Innovation Program
of Chinese Academy of Sciences, and  Director Grants of CASHIPS.
Part of the calculations were performed in Center for
Computational Science of CASHIPS and the Shanghai Supercomputer
Center.

\newpage
%============================== Table 1========================
\begin{table}[tb]
\caption{ The lattice constant in the \emph{c} axial direction,
the electronic ground state (BS), the band gap for semiconductor
and half metal cases, the total supercell magnetic moment (M) per TM and
valence electron distribution (VED) of [TM$_2$(Ant)]$_{\infty}$
(TM= Sc, Ti, V, Cr, Mn and Fe). }

\begin{ruledtabular}
\begin{tabular}{cccccccc}
  meV  &\emph{c}(\AA)    &GS    &band gap (eV)  &M($\mu_{B}$)   &VED  \\
\hline
  Sc   &7.89      &\emph{NN}(metal)        &0.00            &0.00      &3d$^1$4s$^2$  \\
  Ti   &7.37      &\emph{FA}(metal)        &0.00            &0.00      &3d$^2$4s$^2$  \\
  V    &7.01      &\emph{FF}(half metal)   &0.76(direct)       &2.00   &3d$^3$4s$^2$  \\
  Cr   &6.79      &\emph{FF}(half metal)   &1.25(direct)       &0.96   &3d$^4$4s$^2$ \\
  Mn   &6.64      &\emph{NN}(semicond)     &0.66(direct)       &0.00    &3d$^5$4s$^2$    \\
  Fe   &6.87      &\emph{AF}(metal)        &0.00                  &0.00  &3d$^6$4s$^2$      \\
\end{tabular}
\end{ruledtabular}
\end{table}
%==========================  end Table 1 ===========================
\clearpage

\begin{figure}[tp]
\vglue 1.0cm
\newpage
%----------------------------------Fig. 1----------------
\caption{\label{fig:epsart} The top view of
[TM$_2$(Ant)]$_{\infty}$ (a) and five initial magnetic structures used in
our calculations: (b) \emph{NN}, (c) \emph{FF}, (d) \emph{FA}, (E)
\emph{AF}, (f) \emph{AA}. Blue circles and arrows respectively
represent the non-magnetic state and magnetic moment direction of TMs.
Arrows marked by X and Y denote the magnetic interactions inside one layer
and between nearest two layers, respectively. }
%----------------------------------Fig. 2----------------
\caption {Spin-resolved electronic band structures for (a) [V$_2$(Ant)]$_{\infty}$ and (b) [Cr$_2$(Ant)]$_{\infty}$, respectively.
 (c) the total DOS and V 3\emph{d} partial DOS of [V$_2$(Ant)]$_{\infty}$.}

%-----------------------------------Fig. 3---------------

\caption {(a) the spliting of TMs \emph{d} orbitals under the \emph{D$_{6h}$} symmetry crystal
    field. (b) the calculated magnetic moments (per TM ions) of ferromagnetic
[TM$_2$(Ant)]$_{\infty}$ unit cell. The TMs
electrons configurations and magnetic moments predicted by
CT-module are also shown in (b). (c) the total V 3\emph{d} and C
2\emph{p} partial DOS of [V(Bz)]$_{\infty}$. (d) the total V 3\emph{d} and C
2\emph{p} partial DOS of [V$_2$(Ant)]$_{\infty}$.
Schematic HOMO and
LUMO positions of Bz, V and Ant are shown in the left, middle and
right, respectively. The wider blue arrow means a larger
CT between V and Ant.}

%---------------------------------Fig. 4-----------------

\caption { (a) Side view of the relaxed Sc$_4$V$_4$Ant$_5$ cluster.
(b)The schematic structure of branched sandwich clusters
constructed by pentacene, benzene and TMs.}

\end{figure}
\clearpage

%%%%%%%%%%%%%%%%%%%%%%The following is the figures%%%%%%%%%%%%%%%%%%%%%%%%%%%%%%%%%%%%%%%%%%%%%%%%%%%%%%%%%%%%
\newpage
%---------------------------------------Fig. 1----------------------------------------------------
\begin{figure*}[htbp]
\center
{$\Huge\textbf{Fig. 1  \underline{Wang}.eps}$}
\vglue 3.0cm
\includegraphics[width=11.5cm,height=6.5cm,angle=360]{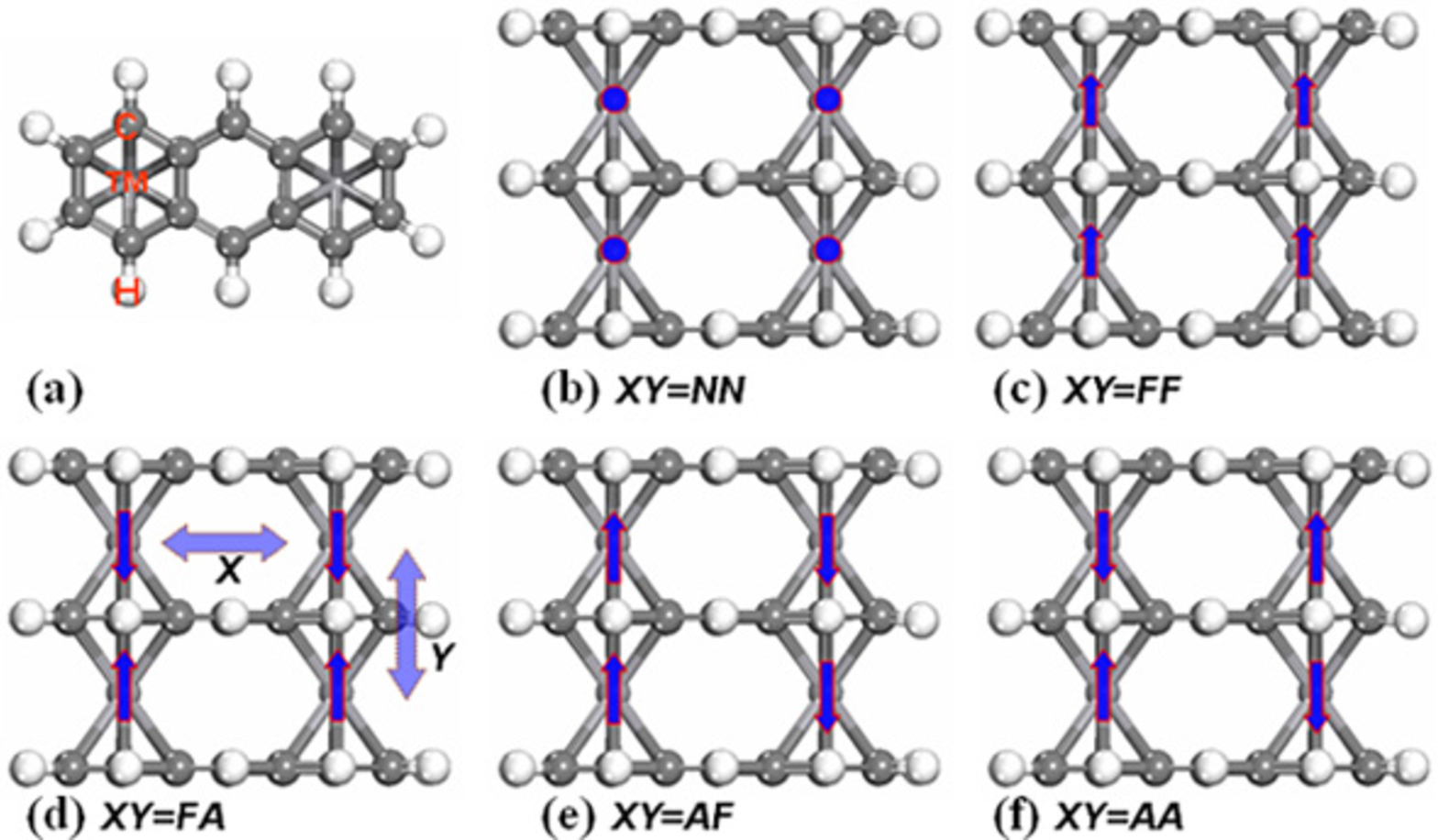}
\end{figure*}
%---------------------------------------Fig. 1----------------------------------------------------
\clearpage
\newpage
%---------------------------------------Fig. 2----------------------------------------------------
\begin{figure*}[htbp]
\center
{$\Huge\textbf{Fig. 2  \underline{Wang}.eps}$}
\vglue 3.0cm
\includegraphics[width=8cm,height=11cm,angle=360]{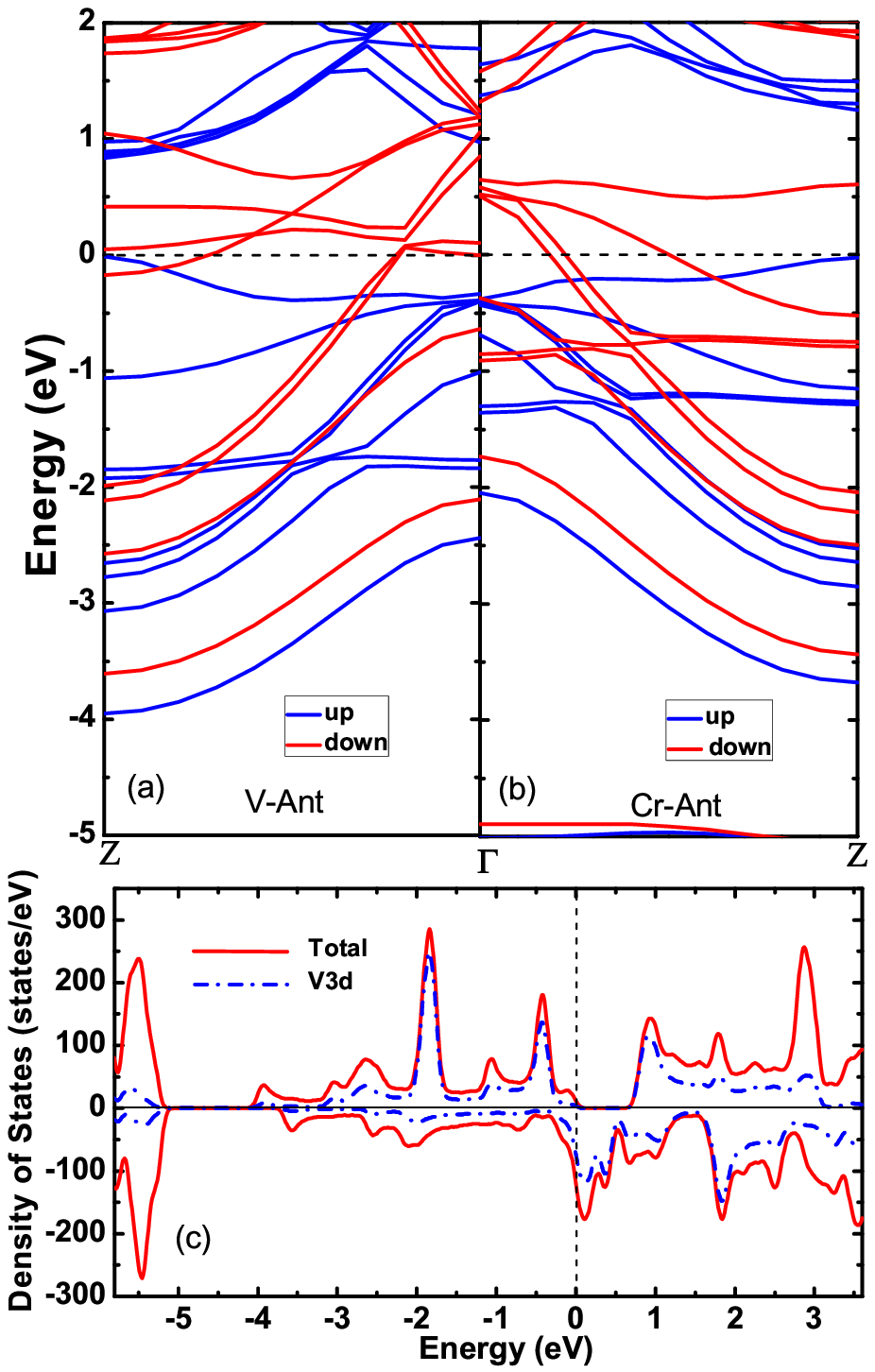}
\end{figure*}
%---------------------------------------Fig. 2----------------------------------------------------
\clearpage
\newpage
%---------------------------------------Fig. 3----------------------------------------------------
\begin{figure*}[htbp]
\center
{$\Huge\textbf{Fig. 3  \underline{Wang}.eps}$}
\vglue 3.0cm
\includegraphics[width=10cm,height=8cm,angle=360]{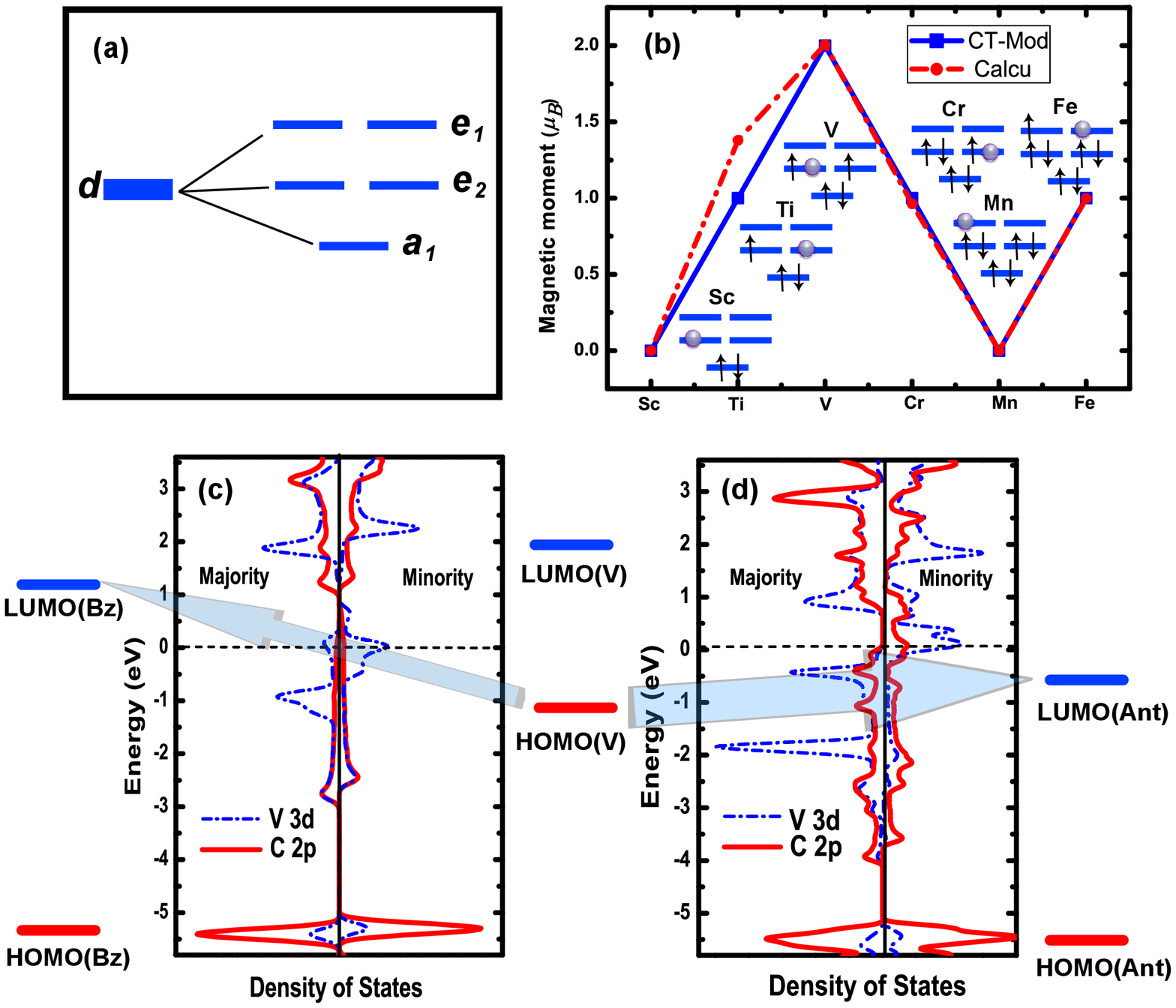}
\end{figure*}
%---------------------------------------Fig. 3----------------------------------------------------
\clearpage
\newpage
%---------------------------------------Fig. 4----------------------------------------------------
\begin{figure*}[htbp]
\center {$\Huge\textbf{Fig. 4  \underline{Wang}.eps}$}
\vglue 3.0cm
\includegraphics[scale=0.8,angle=360]{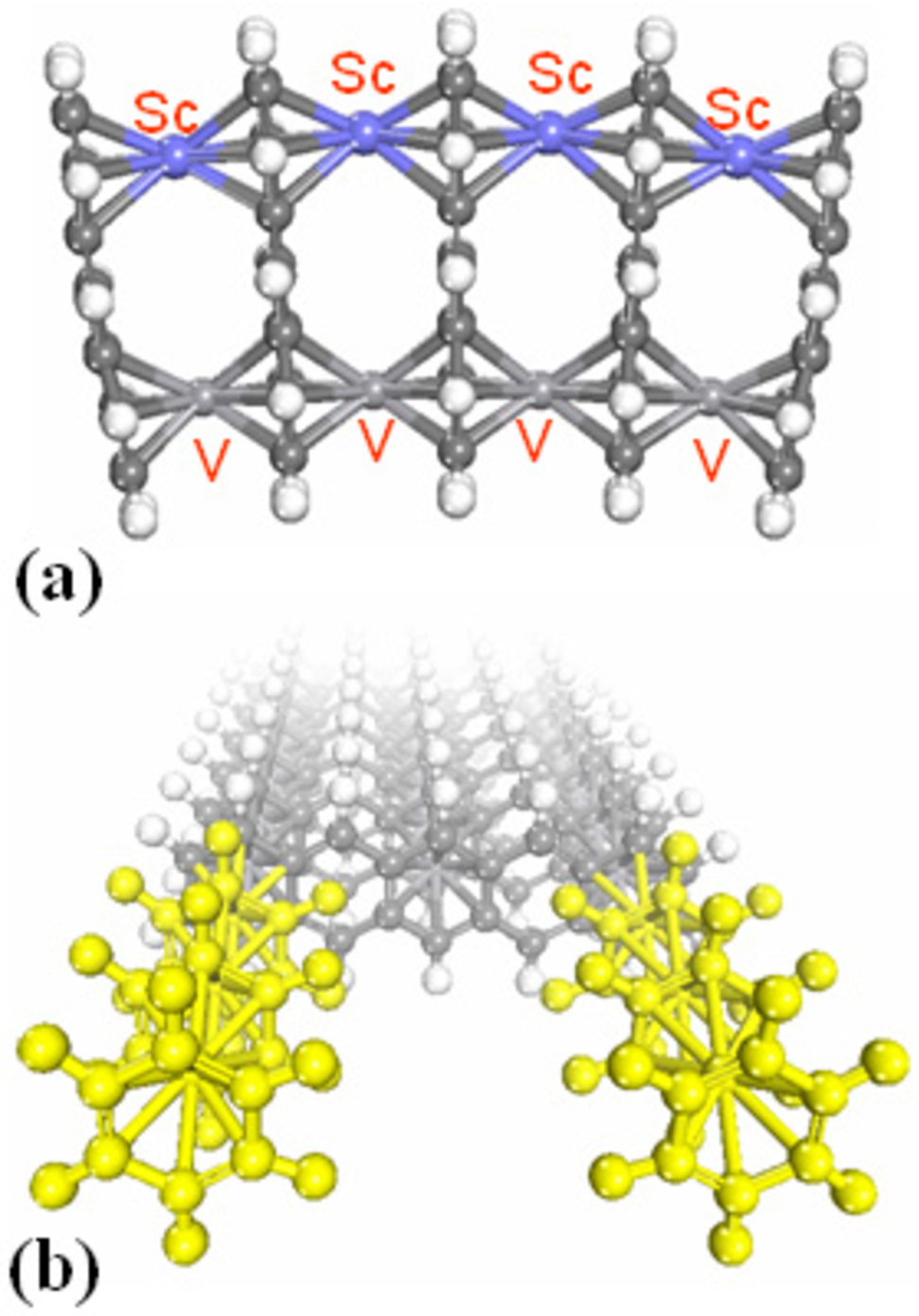}
\end{figure*}
%-------  -------------------------------Fig. 4----------------------------------------------------
\clearpage
\end{document}